\documentclass[aps,pre,showpacs,twocolumn,superscriptaddress,letter]{revtex4}

\usepackage{epsfig}
\usepackage{amsmath,amssymb,wasysym,epsfig,capt-of,ifthen,calc}
\usepackage{latexsym} 
\usepackage{array} 

\usepackage{delarray} 
\usepackage{afterpage} \usepackage{graphicx}
\usepackage{dcolumn}
\usepackage{bm}
\usepackage{float}
\usepackage{longtable}
\newcommand{\be}{\begin{equation}} \newcommand{\ee}{\end{equation}}
\newcommand{\bea}{\begin{eqnarray}} \newcommand{\eea}{\end{eqnarray}}

\usepackage{amsmath}\bibstyle{apsrev}

\begin{document}
\title{Exact solutions for mass-dependent irreversible aggregations}

\author{Seung-Woo Son} \affiliation{Complexity Science Group, University of Calgary, Calgary T2N 1N4, Canada}
\author{Claire Christensen} \affiliation{Complexity Science Group, University of Calgary, Calgary T2N 1N4, Canada}
\author{Golnoosh Bizhani} \affiliation{Complexity Science Group, University of Calgary, Calgary T2N 1N4, Canada}
\author{Peter Grassberger} \affiliation{Complexity Science Group, University of Calgary, Calgary T2N 1N4, Canada} \affiliation{FZ J\"ulich, D-52425 J\"ulich, Germany}
\author{Maya Paczuski} \affiliation{Complexity Science Group, University of Calgary, Calgary T2N 1N4, Canada}

\date{\today}

\begin{abstract}
 We consider the mass-dependent aggregation process $(k+1)X\to
X$, given a fixed number of unit mass particles in the initial
state. One cluster is chosen proportional to its mass and is
merged into one either with $k$-neighbors in one dimension, or --
in the well-mixed case -- with $k$ other clusters picked randomly.
We find the same combinatorial exact solutions for the probability
to find any given configuration of particles on a ring
or line, and in the well-mixed case. 
The mass distribution of a single cluster exhibits scaling laws
and the finite size scaling form is given. The relation to the
classical sum kernel of irreversible aggregation 
is discussed.

\end{abstract}
\pacs{05.70.Ln, 89.75.Da, 89.75.Hc}
\maketitle

Recently the theory of irreversible aggregation was revisited in
view of renormalization of complex networks~\cite{SWSon2010}. In
\cite{SWSon2010}, a simple mapping between random sequential
renormalization (RSR)~\cite{Bizhani2011,Christensen2010} and
irreversible aggregation~\cite{Leyvraz2003} was pointed out, where
a local random renormalization step to produce a new `super-node'
in complex networks corresponds to one aggregation event of
`molecules'. Exact combinatorial analyses, both in one dimension
(without diffusion) and in the well-mixed case, gave the same
scaling law of cluster mass distribution. The corresponding
exponent only depends on $k$, the number of interacting
neighbors~\cite{SWSon2010}. This RSR procedure corresponds to the
`constant' kernel of irreversible aggregation (one of three
well-known ``classical'' kernels-- constant, sum, and product
kernels~\cite{Leyvraz2003}). In this study, we show the relation
between mass-dependent RSR and irreversible aggregation with the
sum kernel. Applying the same combinatorial technique
of~\cite{SWSon2010}, we find the exact solutions for
mass-dependent irreversible aggregation as well.

Here we consider models governed by the reaction $(k+1)X \to X$,
where a cluster is picked randomly, in proportion to its mass,
after which it coalesces with $k$ other clusters. In the case of
one-dimensional models these are $k$-neighbors, while they are $k$
other clusters chosen randomly in the case of well-mixed systems.
In both cases, the other clusters are chosen independent of their
masses. The mass of the newly formed cluster is the sum of the
$(k+1)$ masses. For one dimensional models, both a ring with
periodic boundary condition and a line with open boundary
condition are considered. Reactions are allowed  only if there is
a sufficient number, $k$, of available clusters.

\begin{figure}[b]
\includegraphics[width=0.7\columnwidth]{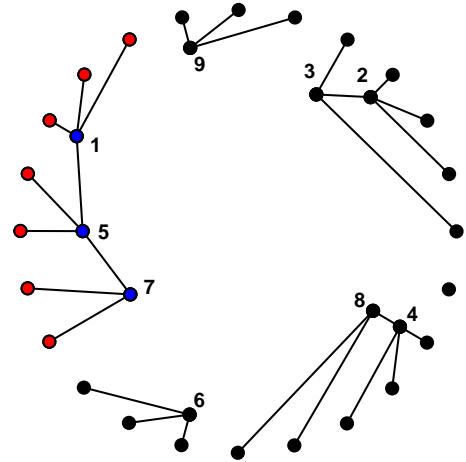}
\caption{(Color online) Illustration of aggregation on a ring with
$k=2$, $N_0=24$, and $N=6$. The tree in color corresponds to a
cluster of mass $m=7$. It has seven leaves (red) and three
internal nodes (blue). The numbers beside internal nodes
correspond to the time when coalescence occurs.} \label{fig0}
\end{figure}

First, let us consider the model defined on a `ring'. Initially,
$N_0$ particles of unit mass ($m=1$) are placed on a ring like
beads (see Fig.~\ref{fig0}). Each particle is labelled by $i\in
[1, ..., N_0]$. At each time, one cluster is picked in proportion
to its mass, and is subsequently merged with its $k$ right
neighbors into one big cluster having a mass equal to the sum of
the $(k+1)$ masses. Cluster masses are therefore restricted to $m
\equiv 1 ~({\rm mod}~k)$. This can be written as $m-1 = ks$, where
$s$ is the number of aggregation events needed to make a cluster
of mass $m$. Similarly the number of clusters at any time, $t$, is
given by $N=N_0-kt$, where time $t$ is denoted by positive
integers representing the total number of aggregation events. We
do not allow two events to happen simultaneously in this study.
Otherwise they can happen either at regular intervals,
intermittently, or according to a Poisson process.

To find the probability that any of the $N$ clusters picked at
random has mass $m$ resulting from $s$ aggregation events, we
follow an approach similar to the one introduced in
Ref.~\cite{SWSon2010}. The crucial observation that makes the
analysis simple is that picking clusters according to their mass
is equivalent to picking {\it sites} with uniform probability,
since a cluster of mass $m$ occupies $m$ sites. Let $i$ be any
site ({\it e.g.} $i=1$), and let $\pi_N^{N_0} (m)$ be the
probability that a cluster of mass $m$ starts at this site and
occupies the sites $(i,i+1,\ldots, i+m-1)$. The probability that
any of the $N$ clusters picked at random has mass $m$ after $t$
events is then \be
    p_N^{N_0} (m) = \frac{N_0}{N} \pi_N^{N_0} (m), \label{eq1}
\ee
and
\be \pi_N^{N_0} (m) =
    \binom{t}{s} \frac{n_{\rm cluster} \times n_{\rm rest}}{ n_{\rm
    total}},
\ee where $n_{\rm cluster}$ is the number of possible histories of
aggregation events $(i_1,i_2,\ldots,i_s)$ leading to a cluster of
mass $m$, $n_{\rm rest}$ is the number of possible ways to form
the other $(N-1)$ clusters, and $n_{\rm total}$ is the total
number of histories for $t$ merging events. The binomial
coefficient ${t}\choose{s}$ corresponds to the number of choices
associated with different time orderings for the $s$ events in the
cluster of mass $m$ and the $(t-s)$ events in the rest of the
clusters.

The total number of all histories involving $t$ events is simply
\be
    n_{\rm total} = N_0^t.                              \label{eq3}
\ee This is to be contrasted to the number of histories $n_N^{[1,
N_0]}$ that lead to the first cluster starting at $i=1$ and the
$N$-th ending at $N_0$. A somewhat more involved argument gives
\be
   n_N^{[1, N_0]} = N \times N_0^{t-1}.            \label{eq4}
\ee The number of histories leading to a single cluster of mass
$m$ covering the sites of interval $[1, m]$ is thus \be
   n_{\rm cluster} = n_1^{[1, m]} = m^{s-1},       \label{eq5}
\ee
while
\be
   n_{\rm rest} = n_{N-1}^{[1, N_0-m]} = (N-1) \times (N_0-m)^{t-s-1}. \label{eq6}
\ee
Combining Eqs.~(\ref{eq1})--(\ref{eq6}), we finally
obtain
\be
    p_N^{N_0} (m) = \frac{N-1}{N} \binom{t}{s}
    \frac{m^{s-1} (N_0-m)^{t-s-1}}{N_0^{t-1}}. \label{pp1}
\ee

For this mass-dependent aggregation process, we can also work out
the joint probability distributions for masses of adjacent
clusters. We denote by $p_N^{N_0} (m_1,m_2)$ the probability to
find a cluster of mass $m_1$ followed immediately to the right by
a cluster of mass $m_2$. This is non-zero only if $m_1=k s_1 +1$
and $m_2 = k s_2 + 1$, where $s_\alpha$ is the number of
aggregation events needed to form a cluster of mass $m_\alpha$. By
the previous arguments, we get \be p_N^{N_0} (m_1, m_2) =
\frac{N-2}{N} \binom{t}{s_0, s_1, s_2} \frac{m_1^{s_1-1}
m_2^{s_2-1} m_0^{s_0-1}}{N_0^{t-1}}, \nonumber \ee where $s_0 =
t-\sum_{\beta=1}^{\alpha} s_{\beta}$ and $m_0 =
N_0-\sum_{\beta=1}^{\alpha} m_{\beta}$. It is interesting to note
that this joint probability properly holds the following relation,
\be p_N^{N_0} (m_1,m_2) = p_N^{N_0} (m_1) p_{N-1}^{N_0-m_1}
(m_2)~. \nonumber \ee

For any $ 1 \leq \alpha \leq N-1$, the joint probability
distribution for $\alpha$ consecutive adjacent clusters is given
by
\be
    p_N^{N_0}
    (m_1,...,m_\alpha)=\frac{N-\alpha}{N}~~ \frac{{\cal
    T}[t,\{s\},\alpha+1] \prod_{\beta=0}^{\alpha} m_\beta^{s_\beta-1}
    }{N_0^{t-1}}, \label{ppj}
\ee
where we used the multinomial coefficient
\be
    {\cal T}[t,\{s\},\alpha+1]= \left(\begin{array}{c} t \\
    s_0,\ldots, s_{\alpha} \end{array}\right) . \nonumber
\ee In particular, this can be done for the joint distribution for
all $N$ masses by setting $\alpha = N-1$. The resulting expression
is always invariant under any {\it permutations} of $N$ numbers
$(m_1,...,m_N)$, as was the case with mass-independent
aggregation~\cite{SWSon2010}. Hence the $N$-cluster probability is
independent of the spatial ordering of the clusters. Therefore,
there are {\em no spatial correlations}, even though there are
obvious correlations between the masses at any given time. For
this reason (and as verified in detail using Eq.~(\ref{eq4})
instead of Eq.~(\ref{eq3})), the joint probability for $N$ masses
on a line, {\it i.e.}, a one-dimensional system with open
boundaries, is also given as Eq. (\ref{ppj}), showing that the two
models lead to the same statistics for any $\alpha$.

The absence of spatial correlations  indicates that the same
dynamics might also result from the well-mixed case. To check
this, we now start with a bucket containing $N_0$ balls, each of
unit mass. An event consists of first picking one ball with
probability proportional to its mass and then choosing $k$ balls
out of the bucket, independent of their masses. The balls are
merged and a new ball, having a mass equal the sum of the masses
of its $(k+1)$ constituents is returned to the bucket. This
process repeats until $N$ clusters remain.

The single cluster mass distribution for the well-mixed model can
be obtained using the same strategy as before. Since events now
correspond to choosing one ball with a mass-weighted probability,
and $k$ balls out of $(N_0-kt-1)$ balls randomly, we have a
$t$-power of $N_0$ and a product of binomial coefficients, \bea
   n_{\rm total} &=& N_0 {N_0-1\choose k}\times \ldots N_0{N+k-1\choose
   k} \nonumber \\ &=& \frac{N_0^t}{(k!)^t} \frac{(N_0-1)!}{(N-1)!} = \frac{N_0^{t-1}}{(k!)^t} \frac{N_0!}{(N-1)!}~.
\eea The expressions for $n_{\rm cluster}$ and $n_{\rm rest}$ are
analogously \be n_{\rm cluster}=\frac{m^{s-1}}{(k!)^s} m! ~~~, \ee
\be n_{\rm rest}=\frac{(N_0-m)^{t-s-1}}{(k!)^{t-s}}
\frac{(N_0-m)!}{(N-2)!} ~. \ee The number of time orderings is
exactly the same as before, but the first factor $N_0/N$ in
Eq.~(\ref{eq1}) has to be replaced by ${1\over N}{N_0\choose m}$.
Putting all these considerations together, many cancellations take
place, leading exactly to Eq.~(\ref{pp1}) again. This argument can
be similarly extended to get the full $N$-particle distribution
function, obtaining exactly the same result as Eq. (\ref{ppj}) for
any $k$ and $\alpha$.

\begin{figure}[t]
\includegraphics[width=\columnwidth]{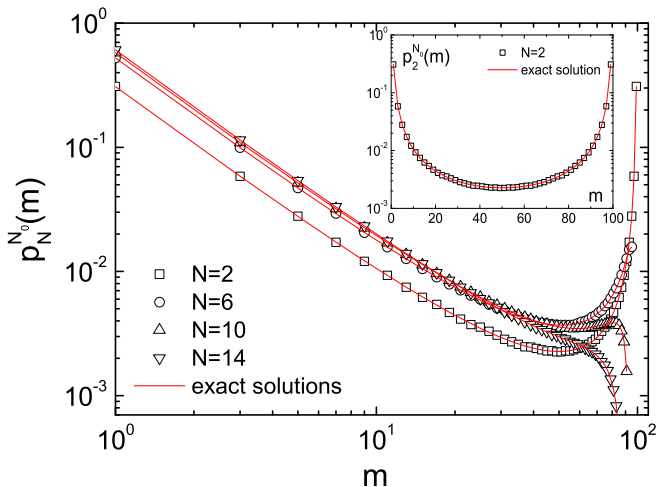}
\caption{(Color online) Cluster size distributions after $t=49$
($N=2$) events for $k=2$, for different values of $N$ averaged
over $10^6$ realizations compared to exact results. The power-law
slope of small $m$ is $-3/2$ independent of $k$.  The large size
behavior changes from an increasing power law to a decreasing one
around $N \sim \sqrt{N_0}$. The inset shows the symmetric
distribution for $N=2$. }\label{fig1}
\end{figure}

Let us look at the characteristics of the solutions. Even though
the composition principle is the same as that of
mass-independent aggregation in Ref.~\cite{SWSon2010}, the final
solution and the characteristics are quite different.
First, cluster size distributions at several different times
are shown in Fig.~\ref{fig1} for $N_0=100$ and $k=2$. The symbols
indicate the numerical simulation results over $10^6$ realizations
and the solid lines are the exact solutions of Eq. (\ref{pp1}).
The tail region corresponding to large cluster sizes changes
from a fast exponential decay to an increasing power law as the
merging process approaches termination. The turning point is
around $N \sim \sqrt{N_0}$. When $N=2$, since the sum of the
two cluster sizes is always $N_0$, the distribution
$p_N^{N_0} (m)$ is symmetric under the exchange $m
\leftrightarrow N_0-m$ for any $k$. The symmetric distribution for $N=2$ is
shown in the inset of the Fig.~\ref{fig1}.

\begin{figure}[b]
\includegraphics[width=\columnwidth]{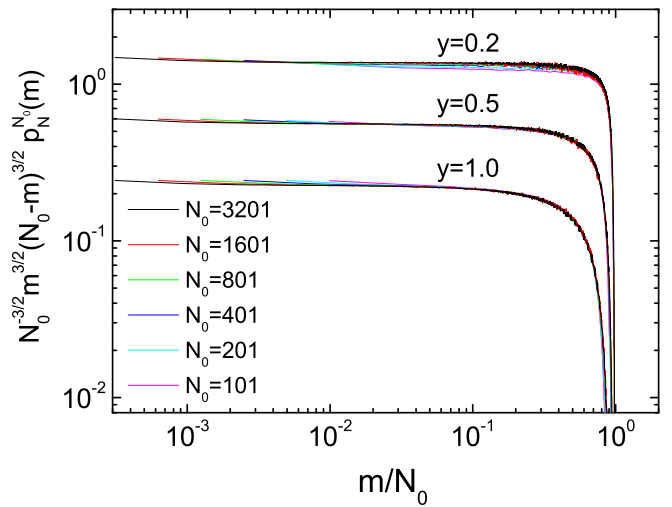}
\caption{(Color online) Finite size scaling collapses for
different $y$ and fixed $k=2$. Collapse lines for $y=0.2$ and
$y=1.0$ are shifted up and down to make them distinguishable from
other collapse lines.}\label{fig2}
\end{figure}
\begin{figure}
\includegraphics[width=\columnwidth]{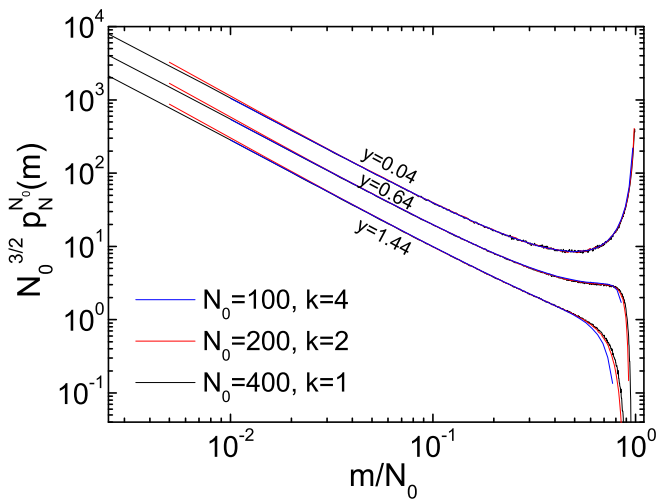}
\caption{(Color online) Finite size scaling collapses for
different $k$. In order to check the scaling collapses for
different $k$, $kN_0$ is fixed at 400 and $N=4$, 16, and 24, which
correspond to $y=0.04$, 0.64, and 1.44. Collapse lines for
$y=0.04$ and $y=1.44$ are shifted up and down to make them
distinguishable from each other.}\label{fig2c}
\end{figure}

When $N_0 \to \infty$, asymptotic power laws can be determined
using Stirling's formula. If $N$ is fixed and both $m$ and
$(N_0-m) \to \infty$,
one obtains the scaling form
\bea
    p_N^{N_0} (m)
    &\sim& N_0^{-\frac{3}{2}} \left[ \frac{m}{N_0} \left(
    1-\frac{m}{N_0} \right) \right]^{-\frac{3}{2}} e^{
    -\frac{N^2}{kN_0}
    \frac{m}{N_0} \left( 1-\frac{m}{N_0}\right)^{-1} }  \nonumber \\
    &\sim& N_0^{-\frac{3}{2}} f\left( \frac{m}{N_0},
    \frac{N}{\sqrt{N_0}} \right). \label{scaling}
\eea
For small masses, this gives a decreasing power law, with exponent $-3/2$,
independent of $k$. Interestingly, this is very different from
mass-independent aggregation, for which the analogous exponent
depends on $k$ and is equal to $-1+1/k$~\cite{SWSon2010}. The exponent $-3/2$
is the same as that for the aggregation with the sum kernel of
the irreversible aggregation obtained in Ref.~\cite{Krapivsky1991}.
Indeed, the rate equation for the current aggregation model in
mean-field theory is the same as for the sum
kernel~\cite{Leyvraz2003,Krapivsky1991}.  For the $k=1$ case, the
rate equation is simply \bea \Delta p_m &=& \sum_{m'=1}^{m} m'
p_{m'} p_{m-m'} - p_m \sum_{m'=1}^{\infty} (m+m') p_{m'} \nonumber
\\ &=& \frac{m}{2} \sum_{m'=1}^{m} p_{m'} p_{m-m'}-(m+\bar{m})p_m,
\label{rateeq} \eea where $p_m$ denotes $p_N^{N_0}(m)$ to make the
equation more concise, and where $\bar{m}$ means the mean cluster
size. Equation ({\ref{rateeq}) is the same as the rate equation
for the sum kernel in~\cite{Leyvraz2003,Krapivsky1991}.
The behavior for large $m$ is different, however, and is
not described by mean field theory.

According to Eq. (\ref{scaling}), $N_0^{\frac{3}{2}} p_N^{N_0}(m)$
should be a function of $m/N_0$ only for fixed \be
  y = \frac{N^2}{kN_0}
\ee and for $N \ll N_0, N_0-m$. The resulting data collapse is
shown in Fig.~\ref{fig2}, where we also factored out a power of
$m/N_0$ to make the curves less steep. Notice that $N$ and $N_0$
are related by $N \equiv N_0 ~({\rm mod}~k)$, which implies that
the values of $y$ used in this plot are not strictly constant but
deviate slightly from their nominal values for small $N_0$, which
causes the deviation from a perfect collapse for $y=0.2$. Even for
different values of $k$, this scaling function works, as can be
seen in Fig.~\ref{fig2c} where the scaling collapses for three
cases, $k=1$, 2, and 4, are shown. Surprisingly, this means that
the process of choosing a cluster proportional to its mass in
conjunction with choosing two clusters at random for $k=2$ is
asymptotically the same as repeating the merging process for $k=1$
twice in the sense of the scaled mass.

The probability $p_N^{N_0} (m)$ satisfies the following recursion
relation
\be
   p_{N+k}^{N_0} (m) =
  A \sideset{}{'}\sum\limits_{m'=m+k}^{N_0-N+1} \frac{m'-1}{m'}
  p^{N_0}_N (m') p^{m'}_{k+1} (m)       \label{rev}
\ee
with
\be
   A = \frac{N_0 N (k+1)}{(N_0-N)(N+k)},
\ee where the prime on the summation symbol indicates that $m'$
must increase in steps of $k$. Interestingly this quadratic
recursion relation corresponds to the time-reversed process of
aggregation, {\it i.e.}, {\em fragmentation}. As with the
quadratic recursion relation of mass-independent
aggregation~\cite{SWSon2010}, the mass distribution at $N+k$ is
given by the product of the mass distribution at $N$ describing
the relative probabilities with which the cluster fragments, given
by $p_{k+1}^{m'}(m)$, and the total fragmentation probability. The
latter was just $\propto (m'-1)$ in the mass-independent
case~\cite{SWSon2010}, while now it is proportional to
$\frac{m'-1}{m'}$. Equation~(\ref{rev}) follows then by
considering how fragmentation leading to a cluster with mass $m$
goes through an intermediary with mass $m'$.

We also examined numerically the aggregation processes where the
clusters were chosen with probabilities proportional to higher
powers of their mass, in particular $\propto m^2$ and $\propto
m^3$, {\it i.e.}, the square of a cluster's mass and the cubic of
a cluster's mass. The asymptotic power law exponents are roughly
about $-5/2$ and $-7/2$ for the $m^2$-dependence and
$m^3$-dependence respectively. However, exact solutions for these
cases have not yet been found.

In summary, we derived the exact solutions for the
probabilities to find any configuration after a fixed number of aggregation
events in the models where a cluster picked with probability proportional
to its mass aggregates with $k$ other particles. More specifically, we
studied three versions of this process (particles on a ring joining with
nearest neighbors, particles on an open-ended line, and the well-mixed case),
and found exactly the same solutions using combinatorial counting. We attribute
this to the absence of spatial correlations, although they are {\it a priori} not excluded.
Differently from the mass-independent random sequential renormalization (RSR),
which shows $k$-dependent exponents in scaling laws for small masses,
the cluster size distribution follows a power-law with exponent $-3/2$ independent of $k$,
which is the same with that of the {\it sum kernel} for irreversible aggregation.
Finally, the aggregation process is also related to a time-reversed fragmentation
process, the characteristics of which are briefly discussed.

 Mass dependent RSR and the related aggregation process was also
considered in two dimensions~\cite{Christensen2010},
 where a runaway giant cluster exists after few
 steps and takes all merging action. The
 behavior is very similar to the {\it gelation} in the
 aggregation process with the product kernel, but aggregation events
in two dimensional RSR involve fluctuating numbers of neighbors,
differently from the aggregation process considered in the present paper.



\end{document}